\documentclass{aa}
\usepackage{graphicx}
\usepackage{natbib}
\def\rxj{\object{RX J1856.5$-$3754}}
\def\Xray{\hbox{X-ray}}

\let\angstrom\AA
\def\AA{\ifmmode{\hbox{\angstrom}}\else\angstrom\fi}
\def\un#1{\ifmmode{\rm\,#1}\else${\rm\,#1}$\fi}

\let\simgt\ga

\let\degree\degr
\def\Sref#1{Sect.~\ref{sec:#1}}
\def\Fref#1{Fig.~\ref{fig:#1}}
\def\Tref#1{Table~\ref{tab:#1}}
\def\apj{ApJ}\def\apjl{ApJ}\def\apjs{ApJS}\def\aap{A\&A}
\def\aj{AJ}\def\an{AN}\def\mnras{MNRAS}\def\nat{Nature}
\def\pasp{PASP}
\begin{document}
\title{Optical spectroscopy and photometry of the neutron star\\
       \rxj\thanks{Based on observations collected at the European
       Southern Observatory, Paranal, Chile (ESO Programmes 63.H-0416
       and 65.H-0643).}}  

\titlerunning{Spectroscopy and photometry of \rxj}

\author{M. H. van Kerkwijk\inst{1}
   \and S. R. Kulkarni\inst{2}}

\institute{Astronomical Institute, Utrecht University,
           P. O. Box 80000, 3508~TA Utrecht, The Netherlands\\
           \email{M.H.vanKerkwijk@astro.uu.nl}
     \and  Palomar Observatory, California Institute of
           Technology 105-24, Pasadena, CA 91125, USA\\
           \email{srk@astro.caltech.edu}
}

\date{Revised version submitted 2001 August 13}

\abstract{We present spectroscopy and imaging with the Very Large
Telescope (VLT) of the neutron star \rxj.  Little is known about the
nature of this source other than that it is a nearby hot neutron star.
Our VLT spectrum does not show any strong emission or absorption
features.  With considerable care to photometric calibration, we
obtain photometric measurements over the optical and ultra-violet (UV)
using our VLT observations and a detailed analysis of archival {\em
Hubble Space Telescope} data.  We find that the entire optical to UV
spectral energy distribution is well described by a slightly reddened
Rayleigh-Jeans tail
($f_\lambda=(2.96\pm0.06)\times10^{-19}(\lambda/5000\un{\AA})^{-4}
10^{-0.4(0.12\pm0.05)(A_\lambda/A_V-1.138)}
\un{erg}\un{s^{-1}}\un{cm^{-2}}\un{\AA^{-1}}$, where $A_\lambda/A_V$
is the reddening curve; implied $V=25.58\pm0.02$).  The reddening is
consistent with the interstellar absorption inferred from X-ray
spectroscopy.  The simplest explanation for this Rayleigh-Jeans
emission is that the optical-UV radiation arises from thermal emission
from the surface of the neutron star.  The high degree to which the
data conform to the Rayleigh-Jeans tail significantly limits
contributions from other sources of emission.  In particular, our
observations are inconsistent with the presence of an accretion disk
and also strongly constrain the amount of magnetospheric emission from
this enigmatic neutron star.
}

\maketitle

\section{Introduction}\label{sec:intro}

The soft \Xray\ source \rxj\ is the brightest and nearest of the
so-called isolated neutron stars\footnote{Given that most radio
pulsars are also isolated, a better name might be `thermally emitting'
neutron stars.} (for a review, see \citealt{ttz+00}).  These objects
have \Xray\ spectra which appear to be entirely thermal, indicating
that the emission arises from the photosphere and that there is little
if any contamination from ill-understood emission processes such as
those occurring in magnetospheres in radio pulsars and accretion flows
in X-ray binaries.  Therefore, these objects offer perhaps the best
hope of modeling neutron-star spectra, and inferring the effective
temperature, surface gravity, and gravitational redshift.  In
principle, this could lead to unique constraints on the equation of
state of matter in the neutron star interiors (e.g., \citealt{lp01}).

Given these possible gains, ever since its discovery in
\citeyear{wwn96} by \citeauthor{wwn96}, \rxj\ has been the subject of
much observational attention.  \citet{wm97} used the {\em Hubble Space
Telescope} ({\em HST\/}) to discover a very faint, $V\simeq25.6$,
optical counterpart; its flux is roughly consistent with seeing the
Rayleigh-Jeans tail of the $kT\simeq50\un{eV}$ spectrum.  Further {\em
HST} observations were used by \citet{wal01} to measure the parallax,
while \citet{pwl+01} used {\em HST}, {\em ROSAT}, {\em ASCA}, and {\em
EUVE} to measure the broad-band spectral energy distribution.

\citet{pwl+01} also presented detailed model atmospheres for a variety
of compositions, with which they were able to model the broad-band
spectrum satisfactorily.  This leads to strong constraints on the
temperature.  When combined with the parallax, however, the inferred
radii are too small for realistic neutron star models.
\citeauthor{pwl+01} suggest the surface may not have a uniform
temperature distribution.  If so, the broad-band spectrum can be used
to set only weak constraints on the equation of state.

If one could observe spectral features in the spectrum, one might be
able to measure the surface gravity and gravitational redshift without
much ambiguity (\citealt{pae97}).  In this respect, the first spectrum
at good resolution, taken with {\em XMM} of RX~J0720.4$-$3125, was
disappointing, as no features were seen (\citealt{pmm+01}).  Recent
results on \rxj\ appear similarly disappointing, with neither X-ray
spectra taken with {\em Chandra} (\citealt{bzn+01}), nor ultra-violet
spectra taken with {\em HST} (\citealt{pwl+01}) showing strong
features.

The use of \rxj\ to address the fundamental issues in physics and
astrophysics mentioned above would benefit from -- or even require --
understanding the nature of the source.  \citet{wwn96} suggested it
could be a young, cooling neutron star, or a neutron star kept hot by
accretion from the interstellar medium.  An alternative would be that
it is a few million-year old magnetar, as was suggested for
\object{RX~J0720.4$-$3125} on the basis of its 8.4-s periodicity
(\citealt{kvk98}).  An indication that \rxj\ might be young is its
proper motion, which \citet{wal01} found to point away from the nearby
Sco-Cen association.  This has led him to the plausible suggestion
that \rxj\ was born in this association about a million years ago.
For a young neutron star, however, it is hard to understand the lack
of X-ray pulsations.  Could it be that this neutron star has no
significant magnetic field?  Almost certainly, the interpretation of
the high resolution X-ray spectra of this source will depend on
knowing the composition of the atmosphere, the strength of the
magnetic field, and the level of non-thermal emission.

\begin{table}
\caption[]{Log of the VLT spectroscopic observations.} 
\label{tab:log99}
\begin{tabular*}{\hsize}{@{}ll@{~}llrll@{}}
\hline\relax\\[-2ex]
Object$^{\rm a}$&\multicolumn{2}{l}{UT$^{\rm b}$}&
Slit or&\multicolumn{1}{c}{$t_{\rm int}$}& Par.$^{\rm c}$&$\sec z$\\
&\multicolumn{2}{l}{day, time}&
Filter&\multicolumn{1}{c}{(s)}&(\degree)\\[.3ex]
\hline\relax\\[-2ex]
\object{EG 274}&15,&23 02&5\arcsec&  $3\times10$\rlap{$^{\rm d}$}&$-76$& 1.21\\
X-F/$+84\fdg6$    &&23 58&B             &         300&            $-94$& 1.67\\
               &16,&00 08&1\arcsec&        2700&                  $-93$& 1.60\\
                  &&00 58&B             & $3\times300$&           $-85$& 1.33\\
                  &&01 17&1\arcsec&$2\times2700$&                 $-82$& 1.26\\
                  &&02 54&5\arcsec&         300&                  $-53$& 1.06\\
X-L/$-8\fdg5$     &&03 11&1\arcsec&$2\times2700$&                 $-44$& 1.05\\
                  &&04 45&5\arcsec&         300&                  $+37$& 1.04\\
X-F/$-95\fdg4$    &&05 00&1\arcsec&     $\ldots$\rlap{$^{\rm e}$}\\
X-F/$-95\fdg4$    &&06 06&1\arcsec&$3\times2100$&                 $+72$& 1.15\\
                  &&07 56&5\arcsec&         300&                  $+93$& 1.57\\
\object{BPM 16274}&&08 08&5\arcsec&  $3\times20$\rlap{$^{\rm d}$}&$-47$& 1.21\\
           &&10 36&5\arcsec&  $3\times20$\rlap{$^{\rm d}$}&$+19$& 1.14\\[.3ex]
\hline\relax\\[-2ex]
\end{tabular*}\\
$^{\rm a}$ X-F and X-L are exposures taken at the indicated position
angle, for which the slit passes over \rxj\ and star F or L (see
\Fref{super});  standards were taken at position
angle~$0^\circ$.\\
$^{\rm b}$ UT day in July 1999.\\
$^{\rm c}$ Parallactic angle at start of observation.\\
$^{\rm d}$ Third exposure taken with a GG~435 order separation filter.\\
$^{\rm e}$ Lost because of a glitch in telescope pointing.
\end{table}

In an effort to understand the nature of this important but enigmatic
source, we have undertaken a series of observations.  In this paper,
we report on the first optical spectrum of \rxj\ and on accurate
optical-UV photometry.  In our spectra, we find evidence for a nebula
around \rxj.  Those observations and the interpretation of the nebula
will be the subject of the next paper (Van Kerkwijk \& Kulkarni, in
preparation).

The organisation of this paper is as follows.  We describe our
observations in Sects.~\ref{sec:spectroscopy} and~\ref{sec:imaging},
and the reduction in \Sref{reduction}.  We pay particular attention to
accurate calibration, since some of our results turn out to be
discrepant with previously published ground-based results.  We also
re-analyse the {\em HST} imaging, taking particular care to correct
for systematic effects affecting faint stars.  We discuss the
results in Sects.~\ref{sec:sed} and~\ref{sec:nonrj}.

\section{Spectroscopic observations}\label{sec:spectroscopy}

We observed \rxj\ on the night of 1999 July 15 to 16 at the 8-m Unit
Telescope \#1 (Antu) of the Very Large Telescope at Paranal, using the
Focal Reducer/Low Dispersion Spectrograph FORS1 to obtain spectra
covering the optical range and a number of images.  A log of the
observations is given in \Tref{log99}.  The conditions were good, with
the seeing varying from $0\farcs8$ to $1\farcs2$.  The night started
with patchy high cirrus, which disappeared later.  The spectroscopy
was done using a $300\un{lines}\un{mm^{-1}}$ grism, the standard
resolution collimator (final f-ratio of 3.13), and a Tektronix CCD
detector with $2048\times2048$ pixels of $24\un{\mu{}m}$.  With this
setup, the plate scale is $0\farcs2\un{pixel^{-1}}$, and the
3600--9000\un{\AA} wavelength range is covered at
$2.6\un{\AA}\un{pixel^{-1}}$.  With the 1\arcsec\ slit, the wavelength
resolution is $\sim\!13\un{\AA}$.  The detector was read out through
all four amplifiers, using the high gain setting
($\sim\!1.6\un{e^-}\un{ADU^{-1}}$) for the spectra, and the low gain
setting ($\sim\!3\un{e^-}\un{ADU^{-1}}$) for the images.

\begin{table}
\caption[]{Log of the VLT imaging observations.}
\label{tab:log00}
\begin{tabular*}{\hsize}{@{}ll@{~}llll}
\hline\relax\\[-2ex]
Object$^{\rm a}$&\multicolumn{2}{l}{UT$^{\rm b}$}&Sequence$^{\rm c}$&
Seeing&$\sec z$\\
                &\multicolumn{2}{l}{day, time}&&(\arcsec)\\[.3ex]
\hline\relax\\[-2ex]
RX J1856& 1,& 06:57--07:52& BR$\alpha$R$\alpha$R& $0.77$&1.14--1.06\\
        &   & 07:54--08:48& BR$\alpha$R$\alpha$R& $0.69$&1.05--1.02\\
RX J1856& 2,& 07:23--08:16& BR$\alpha$R$\alpha$R& $0.47$&1.09--1.04\\
        &   & 08:19--09:11& BR$\alpha$R$\alpha$R& $0.57$&1.03--1.02\\
        &   & 09:14--09:45& BR$\alpha$R&  $0.77^{\rm d}$&1.03--1.04\\
PG 0942 & 2,& 23:10--23:20& BRBR&                 $1.2$& 1.08\\
RX J1856& 3,& 06:00--06:54& BR$\alpha$R$\alpha$R& $0.57$&1.27--1.13\\
        &   & 06:59--07:51& BR$\alpha$R$\alpha$R& $0.64$&1.12--1.05\\
        &   & 08:00--08:52& BR$\alpha$R$\alpha$R& $0.62$&1.04--1.02\\
PG 1657 &   & 09:26--09:51& BRBR&                 $1.3$& 1.60--1.69\\
        &   & 09:54--10:05& BRBR&                 $1.0$& 1.71--1.81\\
RX J1856& 4,& 06:25--07:18& BR$\alpha$R$\alpha$R& $0.75$&1.19--1.09\\
        &   & 07:23--08:16& BR$\alpha$R$\alpha$R& $0.75$&1.08--1.03\\[.3ex]
\hline\relax\\[-2ex]
\end{tabular*}\\
$^{\rm a}$ In full: \rxj, \object{PG 0942$-$029} and
\object{PG 1657+078}.\\
$^{\rm b}$ UT day in May 2000.\\
$^{\rm c}$ For \rxj, the integration times were 138~s in B, 135~s in
R, and 1019~s in H$\alpha$; for the standards, 3 and 20~s in B, and 1
and 20~s in R.\\
$^{\rm d}$ The B-band image has $1\farcs3$ seeing.
\end{table}

\begin{figure*}
\sidecaption{\includegraphics[width=12cm]{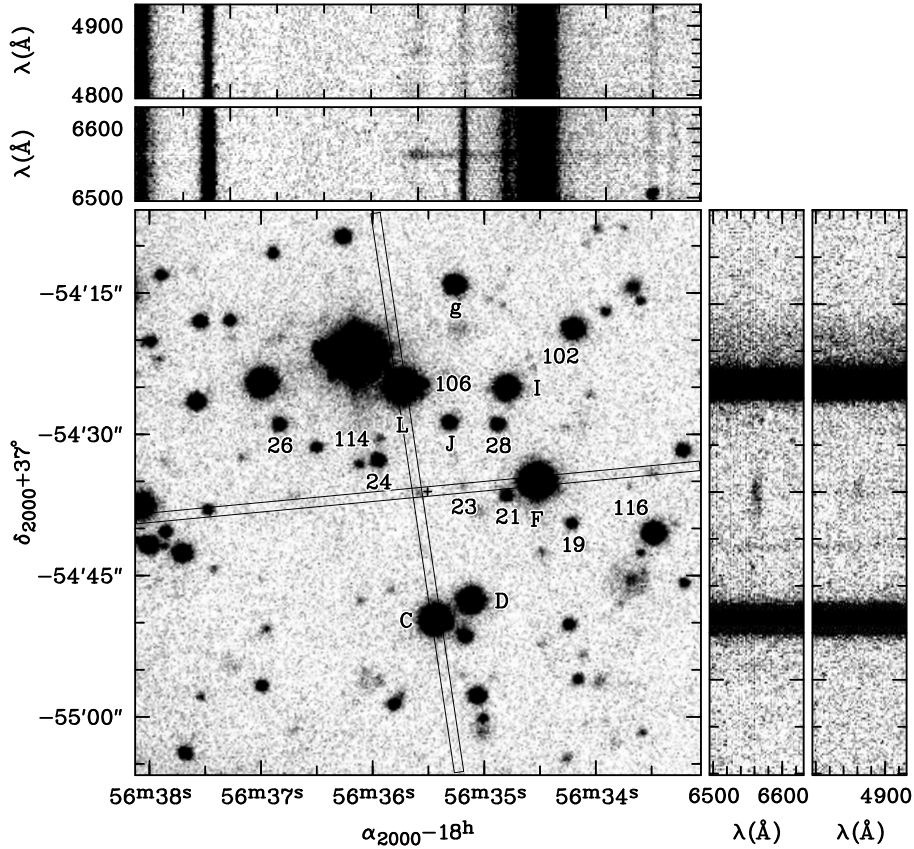}}
\caption[]{Direct and spectral images of \rxj.  The main panel shows
the B-band image obtained at the start of our 1999 spectroscopic run
(\Tref{log99}).  The positions of the two settings of the slit for the
spectroscopy are indicated by the two sets of parallel lines.  The
counterpart to \rxj\ (star X) is within the intersection of the two
slit positions.  The position it had in October 1996, when it was
observed with {\em HST} (\citealt{wm97}), is indicated by the plus
sign.  Previously identified stars for which we obtained photometry
(see \Tref{photometry}) are labeled below their image, except for
stars C and 114 (label on the left) and stars D, I, 102, and 106
(label on the right).  The panels above and to the right of the main
panel show the part around H$\alpha$ and H$\beta$ of the stacked
spectral images for the two slit positions, over stars F and~L,
respectively.  Along the spatial axes, the tick marks indicate
5\arcsec\ intervals along the slit.  The emission from the nebula
around \rxj\ can be seen clearly.  The flux from \rxj\ itself cannot
be seen near~H$\alpha$.}\label{fig:super}
\end{figure*}

Since the source is faint, we set up using two brighter stars nearby,
stars F and~L (here and below we follow the nomenclature of
\citeauthor{wwn96} [\citeyear{wwn96}]; see also \Fref{super}), chosing
the position angle such that the counterpart of \rxj\ (hereafter
star~X) should fall in the slit as well.  To measure the positions, we
reanalysed the images of \citet{wm97}, taken through F606W and F300W
filters with the Wide Field Planetary Camera~2 (WFPC2) on board {\em
HST}.  The analysis of the {\em HST} images is described in more
detail in \Sref{hst}.

In order to minimise slit losses due to differential refraction, we
took set-up images through a Bessel~B filter, used the differential
refraction corrector of FORS1, and chose as reference the star for
which the position angle was closest to the parallactic one.  We
obtained spectra with a total exposure time of 6.5 hour, taking
individual exposures at various positions along the slit in order to
mitigate the effect of bad pixels and other defects.  To calibrate the
slit losses, we followed our sequences of long integrations with short
exposures through a wide slit (5\arcsec, formed using the multi-object
slitlets).  We observed two spectrophotometric flux standards through
the wide slit to calibrate the instrumental response.

For verification of our set-up, we took a 5-minute B-band image before
starting the first spectroscopic observation (which used star~F as
reference).  From the image, star~X appeared to have moved to the
East\rlap{.}\footnote{In the meantime, the proper motion has been
measured accurately using {\em HST} (\citealt{wal01}).}  After the
first spectrum, therefore, we took three additional 5-minute B-band
images\footnote{Photometry of star~X using these images was presented
by \citet{neu01}.  He finds $B=25.1\pm0.4$, consistent with (though
substantially less accurate than) our result.}  to measure a more
accurate position, and found that star~X had, rather fortuitously,
moved exactly along the slit; hence, the first spectrum was not lost.
For star~F, we therefore kept the same set-up, while we changed the
position angle for star~L appropriately.

\section{B, R, and H$\alpha$ imaging}
\label{sec:imaging}

Images in B, R, and H$\alpha$ of \rxj\ and its environment were taken
for us in the nights of 2000 April 30, May 1, May 2, and May 3, at
Unit Telescope \#2 (Kueyen) using the Focal Reducer/Low Dispersion
Spectrograph FORS2.  The standard resolution collimator was used, and
a Tektronix CCD detector with $2048\times2048$ pixels of
$24\un{\mu m}$.  The corresponding plate scale is
$0\farcs2\un{pixel^{-1}}$.

During all nights the seeing was good, varying between 0\farcs55 and
0\farcs8.  The nights of 2000 May 1 and May 2 were photometric, the
night of April 30 mostly clear.  Clouds appeared during the second
half of the night of May 3, which includes the time used for \rxj;
from the count rates, however, it appears that the clouds did not
affect the observations.  In total, ten series of images were taken of
\rxj\ (see \Tref{log00}).  Nine were in the order B, R, H$\alpha$, R,
H$\alpha$, R, and one -- used to fill up time -- was shortened to B,
R, H$\alpha$, R.  Of the latter, the B-band image has much worse
seeing and is not used.  The total exposure time is about 21 minutes
in B, 1 hour in R, and 5 hours in H$\alpha$.  During the photometric
night, several standard fields of \citet{lan92} were observed.  No
useful separate H$\alpha$ calibration images were taken.

\section{Reduction}
\label{sec:reduction}

The data were reduced using the Munich Image Data Analysis System
(MIDAS) and procedures running in the MIDAS environment.  From bias
frames taken before and after a given night, the bias level appeared
somewhat variable, both in time and in position on the detector.
However, relative to the levels found from the overscan pixels
(determined separately for the four amplifiers), it remained constant.
For bias subtraction, therefore, we subtracted both the levels from
the overscan regions in individual frames and an average of the
overscan-corrected bias frames (for the appropriate gain setting).
The averages were determined separately for 1999 and 2000.

\subsection{Photometry}

All images were corrected for sensitivity variations using flat fields
constructed from images of the sky taken at dusk and dawn (for the
observations in 2000, only the dawn frames were used, since this
produced much cleaner results).  Averages were formed of the various
series after filtering out cosmic-ray hits, verifying that even for
the best-seeing images no stars were mistakenly affected.

Photometry was done by first determining the offset of instrumental
magnitudes derived from the average frames from those derived from the
images of the first sequence in the night of 2000 May 2, and then
applying a calibration determined from the three standard fields
observed during that night.  We measured instrumental magnitudes using
the DAOphot package (\citealt{ste87}).  We used an iterative
procedure, in which relatively isolated stars were selected and used
to define a point-spread function (PSF), next the PSF was used to fit
all stars and to subtract all but the PSF stars, and then the cleaned
frame was used to determined an improved PSF, etc.  We found that to
model the variations in the PSF over the frame, a second-order
dependence on position was required.  Aperture corrections were
determined from the difference between the fitted magnitudes and
magnitudes measured in 20-pixel (4\arcsec) radius apertures on the PSF
stars in the final frame in which all non-PSF stars had been removed.

The standard fields were analysed in two separate ways.  For deriving
the calibration using the \citet{lan92} photometry, we simply
determined appropriate aperture magnitudes on the reduced frames (if
not overexposed; in practice, we could only use the short frames).  We
inferred extinction coefficients of 0.126 and 0.076 in B and R,
respectively, which are smaller than the typical values of 0.21 and
0.13 listed by ESO.  We did not have sufficient data to measure the
colour terms accurately, although we could confirm that the colour
term for the B band is significant ($-0.025(B-R)$, i.e., the ESO B
band is bluer than Landolt B), while the colour term for the R band is
negligible.  We estimate that the final uncertainty in the zero points
is about 0.02\un{mag}.

We also tried to calibrate our fields using fainter stars, since for
many faint stars in Landolt fields, \citet{ste00} has obtained
calibrated magnitudes from archive observations.  For this purpose, we
analysed the frames using point--spread function fitting as described
above.  Unfortunately, however, while for the field of PG~1657+078,
there are 32 stars with B-band magnitudes and 44 with R, for the field
of PG~0942$-$029 there is only one star.  As a result, we cannot
derive an accurate solution including extinction terms, but only
confirm the solution found using the Landolt photometry.

In \Tref{photometry}, we list photometry for all point-like objects
which are present in the {\em HST} Planetary Camera images taken
through the F606W filter (\citealt{wwn96,wal01}), and which are
detected in both $B$ and $R$.  One word of caution about the brightest
stars, with $R\la18\un{mag}$, which are overexposed in many of the R-band
images.  The magnitudes of these stars have been determined by PSF
fitting to those pixels which were not overexposed, and are therefore
somewhat more uncertain.  We compared the magnitudes for these stars
with magnitudes inferred from the first series of images, in which
overexposure is less of an issue because of the relatively bad seeing.
We found that the photometry in \Tref{photometry} may slightly
underestimate the true brightness of the $R\la18\un{mag}$ stars, by
$\sim\!0.01\un{mag}$.

\subsection{Spectroscopy}

For the spectroscopy, the flat fielding turned out to be problematic,
because the spectroscopic flats, taken with the internal flat-field
lamp illuminating the instrument cover, had an illumination pattern so
different from that of the actual observations on sky that they were
useless.  Since the observations of the flux standards indicated that
fringing was not a problem and that pixel-to-pixel sensitivity
variations were much smaller than the sky-subtraction uncertainties
for our very faint source, we decided to forego flat-fielding
altogether.  In order to equalise the four quadrants of the chip,
which are read out through amplifiers with slightly different gain, we
multiplied with the gains for the different amplifiers as measured by
the instrument team.  This provided very satisfactory results.

For the sky subtraction, clean regions along the slit within about 100
pixels of the sources of interest were selected, and these were fitted
using a polynomial function.  The order of the polynomial was mostly
zero, but could be increased up to quadratic at any given column as
long as further terms increased the goodness of the fit to the sky
regions significantly.  For each set of observations, the
sky-subtracted images were registered and added together.

From both the individual and the summed sky-subtracted images, spectra
were extracted using an optimal weighting scheme similar to that of
\citet{hor86}.  For this purpose, the spatial profile of the
bright star (either F or~L) was determined, and this was used to
extract optimally weighted spectra at the position of the bright star
itself, as well as at the position of star~X.  Furthermore, for
verification, spectra were extracted at a number of empty positions.
These were all consistent with zero flux. 

The dispersion relation was found using an exposure of helium, argon
and mercury lamps.  Line positions were determined for positions along
the whole slit.  At a single position, a fourth order fit was
sufficient, giving root-mean-square residuals of 0.5\un{\AA}; to obtain
the same residuals for a two-dimensional relation required terms up to
fifth order along the dispersion direction and second order along the
spatial direction (for a total of 18 terms).  The latter solution was
used to calculate the wavelengths for all extracted spectra.

\subsection{Spectral Images}

In the extracted spectra of \rxj, emission lines of H$\alpha$ and
H$\beta$ appeared.  Inspection of the sky-subtracted frames showed
that these lines were extended, especially along the slit over star~F.
In fact, it extended into the regions used to define the sky emission,
and hence it had been partly removed in the sky-subtraction stage.  In
order to provide a cleaner picture, we rebinned the raw images to
spectral images, in which every column is at a constant wavelength,
removed cosmic rays, and formed averages for the two slit positions
(excluding the first spectrum, which had a cosmic-ray hit at H$\beta$
near the target).  Next, we determined the sky emission as a function
of wavelength in regions far away from the neutron star, and
subtracted this from all columns.  The parts of the images around
H$\alpha$ and H$\beta$ are shown in \Fref{super}.  From these spectral
images, it is already clear that the neutron star has a nebula which
is extended along the path it has travelled.  This is confirmed by our
H$\alpha$ imaging.  The H$\alpha$ images and a discussion of the
nature of this nebula will be presented elsewhere (Van Kerkwijk \&
Kulkarni, in preparation).  

\subsection{Flux calibration of the spectra}\label{sec:flux}

For the flux calibration, the spectra were first corrected for
atmospheric extinction using the average La Silla extinction curve.
While this will be only approximately correct, it facilitates the next
step, the determination of the slit losses.  For this purpose, the
ratio with the wide-slit spectra was formed for each of the bright
star spectra taken through the 1\arcsec\ slit.  These ratio spectra
were approximated with second-degree polynomials, which were used to
correct all spectra.

Finally, the spectra were corrected for the response of the
spectrograph derived from the observation of the spectrophotometric
standard \object{EG~274} (\citealt{hws+92,hsh+94}).  The spectra of
this standard were extracted in the same manner as described above,
but in addition a correction was made for the blue second-order light
that overlaps the part of the spectrum at $\simgt\!6500\un{\AA}$ (the
correction was determined with the help of the spectrum taken through
the GG~435 filter).  We observed the DA white dwarf \object{BPM~16274}
as an additional calibrator.  Unfortunately, we realised later that
this star is only calibrated in the ultraviolet.  We still used it to
verify our response curve using a model spectrum kindly provided by
D.~Koester (for $T_{\rm eff}=18750\un{K}$ and $\log{}g=7.83$, as
inferred by \citeauthor{brb95} [\citeyear{brb95}], and normalised to
$V=14.20$, as measured by \citeauthor{egg69} [\citeyear{egg69}]).  For
wavelengths longer than $\sim\!3700\un{\AA}$, the comparison was very
satisfactory, as was a similar comparison using a DA white-dwarf model
provided by D.~Koester for \object{EG~274} ($T_{\rm eff}=24250\un{K}$,
$\log{}g=8.05$; \citealt{vska97}).

The above gives us confidence that the relative calibration over the
spectral range is accurate.  The accuracy of the absolute calibration,
however, is less clear, since the spectra of \object{EG~274} were
taken in the beginning of the night, when some patchy cirrus was still
present.  In order to assess the influence of the cirrus, we compared
fluxes from all B-band (acquisition) images and all wide-slit spectra.
We found that there were variations before about 1\,{\sc{}ut}, but
that after that time the measured count rates indicate the sky was
clear.  To see whether our flux calibration was influenced by the
cirrus, we used the B, V, and R filter curves of \citet{bes90} to
determine synthetic B, V, and R-band magnitudes for all brighter
objects in our wide-slit spectra.  For \object{BPM~16274}, we find
$V=14.20$, $B-V=-0.04$, quite consistent with the observed $V=14.20$,
$B-V=-0.015$ (\citealt{egg69}).  Also for stars~F (twice), L, and C
(in the slit for the star~L position; see \Fref{super}), the synthetic
magnitudes are in good agreement with our B and R-band photometry, as
can be seen in \Tref{flc}.  We conclude that the absolute calibration
of our spectra is accurate to 0.02\un{mag}.

\subsection{Previous photometry of stars L, C, and F}
\label{sec:previous}

While our spectrophotometry of stars L, C, and F agrees well with our
own photometry, it disagrees with measurements in the literature: the
synthetic V-band magnitudes are 0.4\un{mag} brighter than both the
V-band magnitudes of \citet{ntd+97} and the V-band magnitudes inferred
from Gunn g and r measurements of \citet{cms97}.  (The magnitudes of
\citet{wwn96} differ even more, by $\sim\!1\un{mag}$, but
\citeauthor{ntd+97} have already noted that \citeauthor{wwn96} used an
incorrect zero point.)

Comparing colours, we find that our synthetic $B-V$ and $V-R$ colours
are systematically redder and bluer, respectively, than those of
\citet{ntd+97}.  The inferred $B-R$ values, however, are consistent.
This suggests there may be a problem with the V-band only.  Indeed,
the V, $B-V$, and $V-R$ values listed by \citeauthor{ntd+97} imply B
and R-band magnitudes that are in reasonable agreement with our values
for bright stars like stars L, C, and F.  If our synthetic B and R are
correct, however, our synthetic V should be correct too, since the
three bands are tied to each other by {\em relative} calibration on
wide-slit spectra, which proved very reliable on \object{BPM~16274}.

In order to settle the issue, we classified the spectra (see
\Fref{flc}), using the spectral atlases of \citet{sc92} and
\citet{tdw93}.  All three stars appear to be of spectral type~G; see
\Tref{flc}.  For all three, our colours are consistent with the
spectral types (for a small amount of reddening; see \Sref{sed}),
while the $B-V$ and $V-R$ (but not $B-R$) colours of \citet{ntd+97}
and the $g-r$ colour of \citet{cms97} are inconsistent with the
spectral types (independent of reddening).  We thus conclude that,
despite the inconsistencies with earlier work, our calibration is
reliable.

\begin{table}
\caption[]{Spectral types, colours and magnitudes for stars L, C, and
F.  For each star, the first line lists the spectral type inferred
from the spectra and the corresponding intrinsic colours (for a
main-sequence star; \citealt{cox00}).  The following lines list
synthetic magnitudes and colours derived from the wide-slit spectra
(two for star~F), and magnitudes derived from our images.  All
magnitudes and colours have $\sim\!0.02\un{mag}$
uncertainty.}\label{tab:flc}
\begin{tabular}{@{}llll@{}c@{}rr@{}}
\hline\relax\\[-2ex]
Star& Spectral type&
\multicolumn{1}{c}{$B$}&
\multicolumn{1}{c}{$\llap{(}B-V\rlap{)}$}&
\multicolumn{1}{c}{$V$}&
\multicolumn{1}{c}{$\llap{(}V-R\rlap{)}$}&
\multicolumn{1}{c}{$R$}\\
\hline\relax\\[-2ex]
L&G3--6 V--III&& \multicolumn{1}{r}{\null\llap{0.64--0.66}}&&
                 \multicolumn{1}{l}{\null\rlap{0.36--0.37}}\\
 &            &  18.07& 0.74 & 17.33& 0.41& 16.92\\
 &            &  18.06&      &      &     & 16.93\\[.5ex]
C&G6--8 V--III&& \multicolumn{1}{r}{\null\llap{0.66--0.76}}&&
                 \multicolumn{1}{l}{\null\rlap{0.37--0.44}}\\
 &            &  18.56& 0.78 & 17.78& 0.43& 17.35\\
 &            &  18.53&      &      &     & 17.34\\[.5ex]
F&G9--K0 V--III&&\multicolumn{1}{r}{\null\llap{0.79--0.82}}&&
                 \multicolumn{1}{l}{\null\rlap{0.45--0.47}}\\
 &             & 17.94& 0.88& 17.06& 0.46& 16.60\\
 &             & 17.93& 0.86& 17.07& 0.46& 16.61\\
 &             & 17.91&     &      &     & 16.58\\
\hline\relax\\[-2ex]
\end{tabular}
\end{table}

\subsection{Archival HST imaging}\label{sec:hst}

\rxj\ has been a regular target for WFPC2 observations with {\em HST},
both to measure broad-band photometry (\citealt{wm97,pwl+01}) and to
determine the proper motion and parallax (\citealt{wal01}).  We
reanalysed all images (see \Tref{loghst}), both to provide a final
verification of our flux calibration, and to extend the optical
spectral energy distribution for \rxj\ to shorter wavelengths.  In our
analysis, we take into account that during the last years updated zero
points have become available (\citealt{bcgr97}), and that new
prescriptions have been published for correcting for changes in the
amount of contaminants on the CCD windows (\citealt{bg98}), for the
so-called ``long-versus-short anomaly'' (\citealt{cm98}), and for the
slowly degrading charge-transfer efficiency (CTE; \citealt{whc99});
the latter two are particularly important for faint objects.
Furthermore, recently a package specifically written for WFPC2
photometry, {\sc HSTphot}, has been made available by \citet{dol00a},
which includes many of the above corrections (\citealt{dol00b}).

\begin{figure}[t]
\centerline{\includegraphics[width=\hsize]{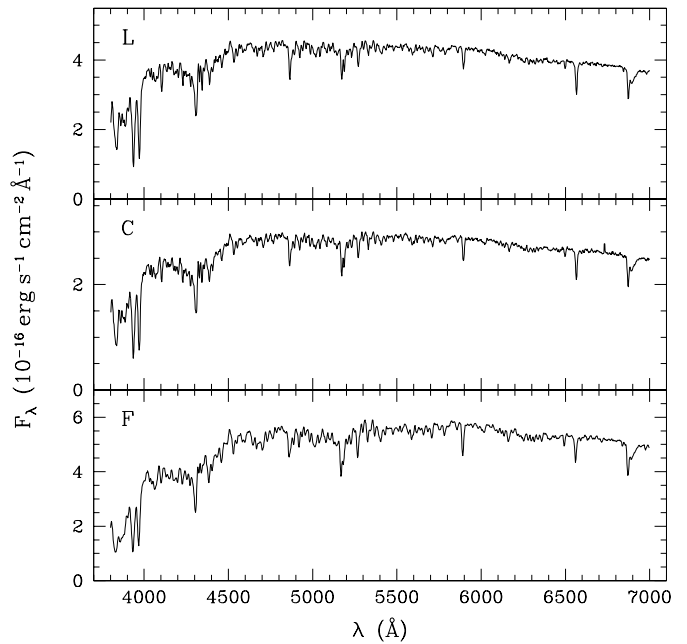}}
\caption[]{Spectra of stars L, C, and F.  These are used for
classification and verification of the flux calibration in
\Sref{previous}; see also \Tref{flc}.}\label{fig:flc}
\end{figure}

Our analysis started with the pipe-line reduced WFPC2 images.  We
measured photometry using {\sc HSTphot}, as well as, for comparison,
our own procedures.  For the {\sc HSTphot} reduction, we followed the
prescription of \citet{dol00a}: (i) mask bad pixels; (ii) combine
images taken at the same position and remove cosmic ray hits; (iii)
determine sky levels; (iv) remove hot pixels; and (v) measure
photometry by point-spread function fitting on the combined image(s).
In the last step, we disabled the determination of point-spread
function residuals and aperture corrections for the F170W image
exposures, since these lack a sufficient number of well-exposed stars.
For the F170W exposures, we also had to make a change to the source
code for the sky determination, viz., to remove the constraint that
the fitted sky level had to be positive.  While this constraint is
physically reasonable, slight inadequacies in the pipe-line
subtraction of the bias and dark current can lead to negative count
rates, which, if not corrected for, lead one to underestimate a
source's brightness; this is indeed the case for the F170W images.
Another change we made to the source code was that we forced the use
of the published (\citealt{dol00b}) charge-transfer efficiency
corrections for all bands (the {\sc multiphot} routine in the {\sc
HSTphot} distribution uses more recent corrections for the optical
filters, but not for the ultraviolet ones; the difference is rather
small).

\begin{table}
\caption[]{Log of archival {\em HST\/}/WFPC2 imaging.  The F606W and
F450W images are taken at two dither positions, different by 5.5
pixels in both X and Y, while the F300W and F170W images are all taken
at the same position}\label{tab:loghst}
\begin{tabular}{@{}lrll@{}}
\hline\relax\\[-2ex]
Identifier&\multicolumn{1}{c}{UT}&
\multicolumn{1}{c}{Filter}&
\multicolumn{1}{c}{$t_{\rm int}$}\\
\hline\relax\\[-2ex]
u3im010[1--4]& 1996 Oct 6&      F606W&$3\times1000+1400$\\
u3im010[5--6]&           &      F300W&$2\times1200$\\
u51g010[1--8]& 1999 Mar 30&     F606W&$8\times900^{\rm a}$\\
u51g030[1--4]&      May 24&     F170W&$2\times2800+2\times2600$\\
u51g040[1--8]&          26&     F450W&$8\times900$\\
u51g020[1--2]&      Sep 16&     F300W&$2\times1300$\\
u51g020[3--6]&            &     F606W&$3\times1300+1290.5$\\[.3ex]
\hline\relax\\[-2ex]
\end{tabular}\\
$^{\rm a}$ u51g0106 has a high background and a large number of
cosmic-ray hits.  It has not been used in the analysis.
\end{table}

In \Tref{photometry}, we list the results.  For the F300W and F606W
filters, the averages of the individual measurements are listed (no
significant variability was found for any star).  Comparing the {\em
HST} photometry with our ground-based results, we find that the two
are consistent.

To verify our technique, we also measured aperture photometry on
averaged images, produced in the manner described by \citet{vkbkk00}.
We measured count rates in apertures with a range of radii, derived
aperture corrections to the standard 0\farcs5 radius aperture,
calculated the corrections discussed above (following the above
references and the {\em{}HST} data and WFPC2 instrument handbooks),
and converted to calibrated magnitudes by applying a 0.1\un{mag}
aperture correction from the 0\farcs5 aperture to infinity and using
the zero points of \citet{bcgr97}.  We note that, as alluded to above,
some corrections are large, especially for faint objects.  For star~X,
the ``long-vs-short'' corrections are $-0.17$, $-0.3$, $-0.3$, and
$-0.4\un{mag}$ in F606W, F450W, F300W, and F170W, respectively (but
see below); the 1996 CTE corrections are $-0.07$ and $-0.13\un{mag}$
in F606W and F300W, respectively; and the 1999 CTE corrections are
$-0.11$, $-0.14$, $-0.30$, and $-0.23\un{mag}$ in F606W, F450W, F300W,
and F170W, respectively.

\begin{table}
\caption[]{Photometry of \rxj\ and stars in the field.  All magnitudes
are in the Vega system.  For star X, the uncertainties are as listed.
For all other stars, an indication of the uncertainty is given by the
presence of no ($\sigma<0.035\un{mag}$), one ($0.035<\sigma<0.075$) and
two colons ($0.075<\sigma<0.125$).  The zero point uncertainty is
about 0.05\un{mag} for the F170W band, and 0.02\un{mag} for all others.}
\label{tab:photometry}
\def\doerro0.#1|{\ifnum#1>3\ifnum#1>7\rlap{::}\else\rlap{:}\fi\fi}
\newcounter{me}
\setcounter{me}{0}
\def\me|#1|#2|{\stepcounter{me}\ifnum\value{me}<7%
\hbox{\vtop{\tabskip=0em\halign{\strut##\hfil\cr
#1\cr\phantom{0}\llap{$\pm$}#2\cr}}}%
\else#1\doerro#2|\fi}
\def\nomagerr{\multicolumn{1}{c}{$\ldots$}}
\def\zb#1{\multicolumn{1}{c}{\hbox to0pt{\hss #1\hss}}}
\begin{tabular*}{\hsize}{@{}lrrrrrr@{}}
\hline\relax\\[-2ex]
Object$^{\rm a}$&
\multicolumn{1}{c@{}}{$m_{\rm 170}$}&
\multicolumn{1}{c}{$m_{\rm 300}$}&
\multicolumn{1}{c}{$B$}&
\multicolumn{1}{c}{$m_{\rm 450}$}&
\multicolumn{1}{c}{$m_{\rm 606}$}&
\multicolumn{1}{c@{}}{$R$}\\[.3ex]
\hline\relax\\[-2ex]
 X  &\me|21.37|0.10|&\me|23.00|0.07|&\me|25.19|0.07|&\me|25.21|0.06|&\me|25.64|0.02|&\me|25.80|0.11|\\
\\[-2ex]
 C  &\me|21.62|0.12|&\me|19.30|0.01|&\me|18.53|0.00|&\me|18.43|0.00|&\nomagerr      &\me|17.34|0.01|\\ 
 D  &\nomagerr      &\me|20.80|0.02|&\me|19.28|0.00|&\me|19.07|0.00|&\nomagerr      &\me|17.69|0.01|\\ 
 F  &\me|20.70|0.09|&\me|19.01|0.01|&\me|17.91|0.00|&\nomagerr      &\nomagerr      &\me|16.58|0.01|\\ 
 I  &\nomagerr      &\me|21.12|0.03|&\me|19.60|0.00|&\me|19.49|0.01|&\nomagerr      &\me|17.89|0.01|\\ 
 J  &\nomagerr      &\me|23.42|0.09|&\me|21.83|0.01|&\me|21.64|0.01|&\me|20.39|0.00|&\me|20.08|0.01|\\ 
 L  &\me|20.93|0.12|&\me|18.72|0.01|&\me|18.06|0.00|&\nomagerr      &\nomagerr      &\me|16.93|0.01|\\ 
 g  &\nomagerr      &\me|20.74|0.03|&\me|20.48|0.00|&\me|20.44|0.01|&\me|19.72|0.00|&\me|19.48|0.01|\\ 
19  &\nomagerr      &\nomagerr      &\me|22.95|0.01|&\me|22.66|0.01|&\me|21.40|0.00|&\me|20.99|0.01|\\ 
21  &\nomagerr      &\nomagerr      &\me|22.60|0.01|&\me|22.25|0.02|&\me|20.48|0.00|&\me|19.97|0.01|\\ 
23  &\nomagerr      &\nomagerr      &\me|25.56|0.12|&\me|24.97|0.05|&\me|23.26|0.01|&\me|22.68|0.01|\\ 
24  &\nomagerr      &\nomagerr      &\me|21.98|0.01|&\me|21.62|0.01|&\me|19.80|0.00|&\me|19.34|0.01|\\ 
26  &\nomagerr      &\nomagerr      &\me|22.28|0.01|&\me|22.19|0.02|&\me|20.90|0.00|&\me|20.56|0.01|\\ 
28  &\nomagerr      &\me|23.10|0.07|&\me|22.16|0.01|&\me|21.93|0.01|&\me|21.05|0.00|&\me|20.79|0.01|\\ 
102 &\nomagerr      &\nomagerr      &\me|25.77|0.11|&\me|25.37|0.08|&\me|23.93|0.01|&\me|23.49|0.02|\\ 
106 &\nomagerr      &\nomagerr      &\me|23.13|0.02|&\me|22.84|0.01|&\me|21.83|0.00|&\me|21.53|0.01|\\ 
114 &\nomagerr      &\nomagerr      &\me|24.66|0.05|&\me|24.88|0.04|&\me|23.88|0.01|&\me|23.31|0.02|\\ 
116 &\nomagerr      &\nomagerr      &\me|25.35|0.09|&\nomagerr      &\me|23.55|0.01|&\me|23.05|0.01|\\[.8ex]
$\Delta_{\rm ST}{}^{\rm b}$&\zb{$-0.412$}&\zb{$-0.024$}&\zb{$-0.602$}&
                            \zb{$-0.475$}&\zb{$+0.316$}&\zb{$+0.555$}\\
$\langle\lambda\rangle^{\rm c}$&\zb{1760}&\zb{2955}&\zb{4371}&
                                \zb{4520}&\zb{5933}&\zb{6464}\\
$\langle A_\lambda/A_V\rangle^{\rm c}$&\zb{2.648}&\zb{1.931}&\zb{1.336}&
                                   \zb{1.297}&\zb{0.949}&\zb{0.841}\\[.3ex] 
\hline\relax\\[-2ex]
\end{tabular*}\\
$^{\rm a}$ X is the optical counterpart.  For the other objects, the
nomenclature follows \citet[][uppercase letters]{wwn96}, \citet[][two
digit numbers]{cms97}, \citet[][three-digit numbers]{wal01} and
ourselves (object g). See \Fref{super}.\\
$^{\rm b}$ Magnitude differences between the ST and Vega systems.
These can be used to calculate the flux by
$f_\lambda=10^{-0.4(m+\Delta_{\rm ST}+21.100)}
\un{erg}\un{s^{-1}}\un{cm^{-2}}\un{\AA^{-1}}$.  From \citet[][WFPC2
filters]{bcgr97} and \citet[][B and R]{bcp98}.\\
$^{\rm c}$ Effective wavelength and reddening for a Rayleigh-Jeans
($f_\lambda\propto\lambda^{-4}$) spectrum.  See
App.~\ref{sec:appendix}.
\end{table}

Comparing the results with those derived using {\sc HSTphot}, we found
good consistency for the brighter stars.  For the fainter stars like
star~X, however, this was only the case if we did not apply the
``long-versus-short'' correction.  Indeed, no such correction is
applied in {\sc HSTphot}, since \citet{dol00b} has not found any
evidence for it; he argues that its appearance likely reflects
inaccurate sky subtraction in the procedures used by \citet{cm98}.  We
do not have strong independent evidence either way, but note that if
we do apply the correction, our VLT photometry for faint objects like
star~X becomes inconsistent with the {\em HST} results (while the
results remain consistent for the brighter stars, since these are not
affected).

Comparing to the magnitudes listed by \citet{wm97} and \citet{wal01},
we find that our results are roughly consistent for the brighter
stars, but that they differ for the fainter stars; in particular, for
star~X, while the F606W magnitude is virtually identical, our F300W
magnitude is 0.4\un{mag} brighter than that of \citeauthor{wm97} (note
that these authors list magnitudes on the ST system).  We suspect that
the differences largely reflect our use of up-to-date corrections.
This suspicion is strengthened by the fact that \citet{pwl+01}, after
making similar corrections, also find fluxes that differ from those of
\citet{wm97}.  Indeed, their results are very similar to ours (their
fluxes are fainter by $0.04\ldots0.10$\un{mag} and their uncertainties
slightly larger, since they use aperture photometry rather than
point-spread function fitting).

\section{Spectrum and spectral energy distribution}
\label{sec:sed}

The reduced, flux-calibrated spectrum is shown in \Fref{optspectrum}.
We limit ourselves to the wavelength range of 3800--7000\un{\AA},
since at shorter wavelengths the flux calibration becomes unreliable,
while at longer wavelengths second-order light starts to contribute
significantly for objects as blue as star~X (it is negligible
shortward of $6500\un{\AA}$ and rises approximately linearly to 10\%
at 7000\un{\AA}).  We also removed 20\un{\AA}-wide regions around
H$\alpha$, H$\beta$, and H$\gamma$, which are (or might be in case of
H$\gamma$) contaminated by nebular emission.  

The spectrum does not contain any significant features.  The best
limits to the equivalent width of any feature are obtained shortward
of $\sim\!5400\un{\AA}$: about 16\un{\AA} for features with 50\% depth
beneath the continuum, and about 60\un{\AA} for features with 25\%
depth.

\begin{figure}
\centerline{\includegraphics[width=0.98148\hsize]{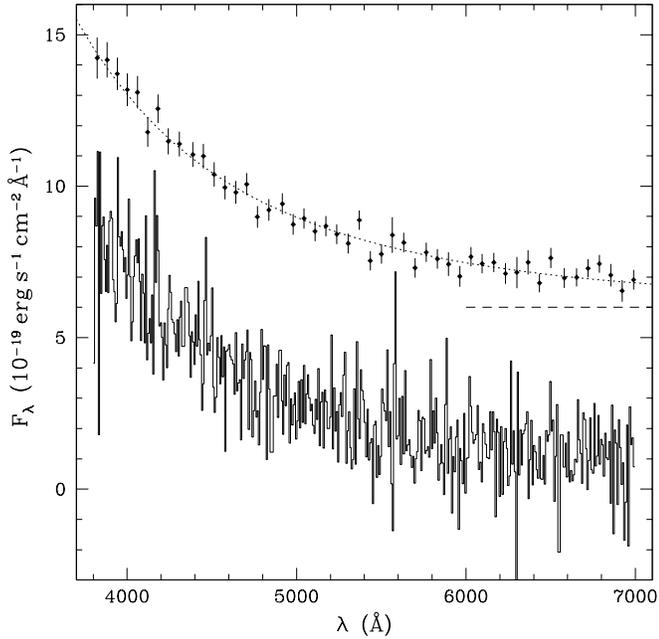}}
\caption[]{Optical spectrum of \rxj.  For the histogram, the data were
averaged in 3-pixel wide bins ($\sim\!7.5\un{\AA}$ or $\sim\!60\%$ of
a resolution element).  The points with (Poisson) error bars reflect
averages over 25-pixel ($\sim\!63\un{\AA}$) wide bins.  These are
offset vertically; their zero level is indicated by the dashed line.
The dotted line indicates the absorbed Rayleigh-Jeans spectrum that
best fits the photometry.  The observed spectrum is consistent with
this and has no significant features.\label{fig:optspectrum}}
\end{figure}

The spectrum is shown together with the photometry in \Fref{optsed}.
One sees that the spectrum is consistent with both the VLT and {\em
HST} photometry.  Indeed, integrating the spectrum over the B and
F450W bandpasses, we infer $B=25.22$ and $m_{450}=25.25$, which
compares well with our photometry (\Tref{photometry}), giving
additional confidence in the calibration of all three data sets.

Both spectrum and photometry indicate a spectral energy distribution
close to that of a Rayleigh-Jeans tail, as would be expected for this
very hot source.  Assuming an intrinsic Rayleigh-Jeans spectrum, we
determine the reddening to the source by fitting a reddened
$\lambda^{-4}$ spectral distribution to the photometry.  We do not
include the spectrum in the fit, because we consider its absolute flux
calibration somewhat less reliable and also because we wish to use it
to verify the result from the photometry.  We use the following
relation between Vega magnitude $m_{\rm Vega}$ and input spectrum
$f_\lambda$,
\begin{equation}
m_{\rm Vega} = -2.5\log 
   \left(\frac{\int f_\lambda\frac{\lambda}{hc}R_\lambda\,{\rm d}\lambda}
              {\int \frac{\lambda}{hc}R_\lambda\,{\rm d}\lambda}
   \right)-21.100-\Delta_{\rm ST},
\label{eq:vegamagdef}
\end{equation}
where $R_\lambda$ is the response at wavelength $\lambda$, and
$\Delta_{\rm ST}$ is the magnitude difference between the Vega and ST
systems (see \Tref{photometry}).  In the latter system, a flat
spectrum has the same magnitude in all filters; it matches the Johnson
system at~V.  The integrations are done over photon rates, since what
we measure with CCDs is the rate of photons in a particular band.  The
response curves we used for the WFPC2 filters are the system response
curves for the planetary camera, taken from the Space Telescope
Science Institute web site.  For the $B$ and $R$ filters, we used the
Landolt filter curves from \citet{bes90}.  For the input spectrum, we
take
\begin{equation}
f_\lambda=f_{\lambda_0}\left(\frac{\lambda}{\lambda_0}\right)^{-4}
          10^{-0.4(A_\lambda-A_{\lambda_0})},
\label{eq:rjspectrum}
\end{equation}
where $f_{\lambda_0}$ is the observed flux at reference wavelength
$\lambda_0$ and $A_\lambda$ is the reddening.  We use the reddening
curve of \citet{ccm89} for $R=3.1$; we include corrections for the
optical as described by \citet{odon94}.  To minimise the covariance
between $f_{\lambda_0}$ and $A_V$, we chose $\lambda_0=5000\un{\AA}$.

\begin{figure}
\centerline{\includegraphics[width=\hsize]{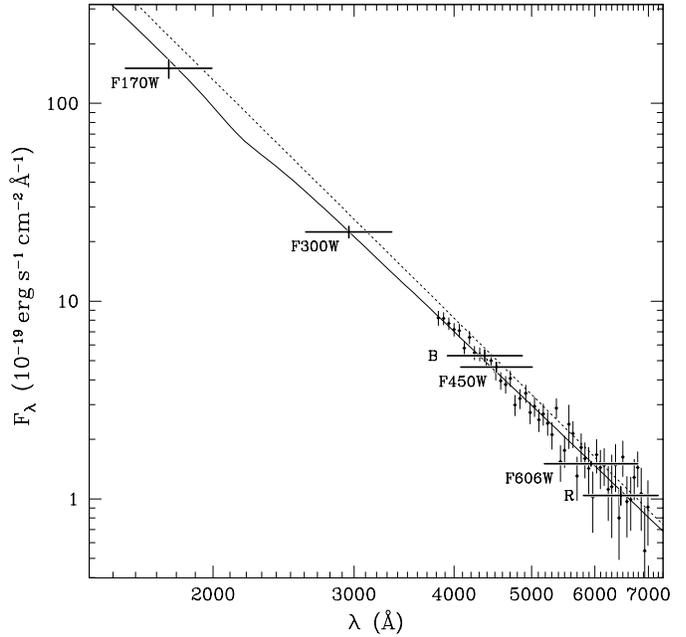}}
\caption[]{Optical/ultraviolet spectral energy distribution of \rxj.
The thick-set points indicate fluxes derived from VLT and {\em HST}
photometry.  The vertical lines indicate the $1\sigma$ errors while
the horizontal ones are measures of the filter widths.  Overdrawn are
the best-fit absorbed (drawn curve) and unabsorbed (dotted curve)
Rayleigh-Jeans spectra, as well as the spectrum observed with the VLT,
averaged in 25-pixel wide bins
($\sim\!63\un{\AA}$).\label{fig:optsed}}
\end{figure}

From the fit to the photometry, we find
$f_{5000}=(2.96\pm0.06)\times10^{-19}
\un{erg}\un{s^{-1}}\un{cm^{-2}}\un{\AA^{-1}}$ and $A_V=0.12\pm0.05$;
the fit is acceptable, with $\chi^2_{\rm red}=3.5$ for four degrees of
freedom (six bands and two parameters; note that for the uncertainties
we used the measurement errors with the zero-point uncertainties added
in quadrature; see \Tref{photometry}).  For the reddening curve used,
$A_{5000}/A_V=1.138$, and thus the unabsorbed flux is
$f_{\lambda_0,0}=(3.36\pm0.17)\times10^{-19}
\un{erg}\un{s^{-1}}\un{cm^{-2}}\un{\AA^{-1}}$, where the error is
dominated by the uncertainty in~$A_V$.  The fit is shown in
\Fref{optsed}; it can be seen that it also is a good fit to the
optical spectrum (see also \Fref{optspectrum}), with $\chi^2=37$ for
49 degrees of freedom (no free parameters; also for other choices of
binning, one finds $\chi^2_{\rm red}\simeq1$).  

The inferred reddening is consistent with the range
$A_V\simeq0.05\ldots0.12$ expected\footnote{We used $A_V=5.6(N_{\rm
H}/10^{22}\un{cm^{-2}})$, as inferred by \citet{ps95} for distant,
highly absorbed objects.  In these, the line of sight samples all
phases of the interstellar medium; hence, the relation may not be
valid locally.}  from the range in X-ray column density $N_{\rm
H}=(1.0\ldots2.2)\times10^{20}\un{cm^{-2}}$ found from different model
fits to the X-ray and EUVE spectrum (\citealt{pwl+01,bzn+01}).  It is
also consistent with the limit set by the total amount of reddening
along this line of sight, which we can infer from stars L, C, and~F.
From the difference between the observed colours and the intrinsic
colours for these stars (inferred from their spectral types; see
\Tref{flc}), we infer $E_{B-R}=0.10\pm0.03$, which corresponds to
$A_V=0.20\pm0.06$.  Using this reddening, and assuming L, C, and F are
main-sequence stars with absolute magnitudes $M_V\simeq5.0$, 5.4, and
5.8 (\citealt{cox00}), respectively, their distances are 2.7, 2.8, and
1.6\un{kpc}, respectively\rlap{.}\footnote{If they are giants, with
$M_V\simeq0.8$ (\citealt{cox00}), they would be at $\sim\!20\un{kpc}$
and might be associated with the Sagittarius dwarf galaxy
(\citealt{igi95}), which is at $\sim\!25\un{kpc}$ and $\sim\!7\degr$
away from \rxj.  Indeed, stars~L and~C have radial velocities of
$+90\pm70\un{km}\un{s^{-1}}$, consistent with being associated.
Star~F, with a radial velocity of $-130\pm70\un{km}\un{s^{-1}}$,
almost certainly is not associated.}  Thus, they are well in the
background relative to star~X, as well as relative to the CrA cloud
complex, which, apparently, contributes very little extinction in this
line of sight, unlike what was suggested previously (\citealt{wwn96}).

\section{Deviations from a Rayleigh-Jeans tail?}\label{sec:nonrj}

Above, we assumed a $\lambda^{-4}$ spectral energy distribution for
star~X.  Indeed, from the observations, we cannot determine the slope
of the spectrum independently: spectra with
$f_\lambda=f_{\lambda_0}(\lambda/\lambda_0)^{-\alpha}$ reproduce the
photometry equally well as long as $A_V=0.12\pm0.06+0.89(\alpha-4)$.
However, stars L, C, and F pose an upper limit of 0.20\un{mag} on the
reddening to star~X.  Using $0<A_V<0.20$, one infers that
$3.8\la\alpha\la4.1$, i.e., the spectrum has to have a slope very
close to that of a Rayleigh-Jeans tail.

\subsection{Limits to a non-thermal component}

Our data place stringent constraints on any contribution from
non-thermal emission.  For instance, fitting a sum of a Rayleigh-Jeans
tail and a non-thermal spectrum with $f_\nu\propto\nu^{0.1}$ (i.e.,
$f_\lambda\propto\lambda^{-2.1}$), such as observed for the
\object{Crab pulsar} (\citealt{sll+00,ccz00}), the
best-fit has zero contribution from non-thermal emission and the same
reddening as inferred above.  Even for zero reddening, the best-fit
contribution is only 11\% at 6000\un{\AA} (for this fit, $\chi^2=5$).
The 95\% confidence upper limit to a non-thermal contribution for a
Crab-like spectrum is 20\% at 6000\un{\AA} (for a non-thermal spectrum
with $f_\nu\propto\nu^{-0.8}$, as observed for \object{Geminga}
[\citealt{mhs98}], this reduces to 8\%).

\subsection{Limits to emission from an accretion disk}

We also considered whether an accretion disk might be present, from
which \rxj\ is accreting.  Accretion from ``debris disks'' has been
invoked in models of anomalous X-ray pulsars (e.g.,
\citealt{vptvdh95,chn00}).  Furthermore, \citet{phn00} suggested that
the deviation from a Rayleigh-Jeans spectrum found from optical
observations of \object{PSR~B0656+14} (\citealt{kpz+01} and references
therein) could be due to the presence of such a disk.  For \rxj, we
considered two cases.  For the first, we assumed the source is powered
by accretion from a disk, in which case both viscous heating
(\citealt{ss73}) and irradiation by the neutron star
(\citealt{vrg+90}) lead to optical emission.  We used routines
described by \citet{hvkk00,hvkvk00} to calculate the emission from
both processes\rlap{,}\footnote{Assuming that the disk is optically
thick and radiates as a black body.  It is questionable whether these
assumptions are justified for the very low accretion rate considered
here, but given the speculative nature of the presence of an accretion
disk, we felt a more detailed analysis was not warranted.}
integrating between an inner radius $r_{\rm in}$ and an outer radius
of $10^{14}\un{cm}$.  For the neutron star spectrum, we take the
black-body fit of \citet{pwl+01} that best fits the observed X-ray to
optical spectral energy distribution ($kT=48\un{eV}$,
$R/d=0.11\un{km}/\un{pc}$) and a distance of 60\un{pc} as inferred
from the parallax (\citealt{wal01}).  We found that for a disk
extending all the way in to the neutron star
($r_{\rm{}in}=10^6\un{cm}$), the optical emission predicted far
exceeds that observed, by three orders of magnitude in~R.  In order
for the emission to remain below 10\% of the R-band flux (i.e.,
$R_{\rm disk}\ga28\un{mag}$), the inner radius had to be
$\ga\!2\times10^9\un{cm}$.  This could be the radius where the disk is
disrupted by a magnetic field; if so, and if the neutron star were
rotating at equilibrium, its period would have to be $\ga\!40\un{s}$.
If such a disk were present, its spectrum would be very red.  At the
limit, one would predict $J=22$ and $K=19\un{mag}$.

In principle, the neutron star could have a disk even if the X-ray
emission is not due to accretion, in which case the mass accretion
rate could be lower and hence the viscous heating of the disk much
reduced.  In the second case we considered, therefore, we ignored the
contribution of viscous heating.  Also for this case, the disk cannot
extend all the way in to the neutron star (it would still exceed the
observed R-band flux by an order of magnitude); we find
$r_{\rm{}in}\ga3\times10^8\un{cm}$ ($\Rightarrow P\ga3\un{s}$; at the
limit, the predicted infrared magnitudes are $J=23$ and $K=20$).

We conclude that all measurements are consistent with a slightly
reddened Rayleigh-Jeans spectrum, with no evidence for features or for
a contribution from non-thermal emission or an accretion disk.

\begin{acknowledgements}
We thank the ESO staff, in particular Thomas Szeifert and Hermann
B\"ohnhardt, for their expert help with both observing runs.  This
research made use of the SIMBAD data base.  The Munich Image Data
Analysis System is developed and maintained by the European Southern
Observatory.  MHvK acknowledges support of a fellowship from the Royal
Netherlands Academy of Science, and SRK support from NASA and NSF.
\end{acknowledgements}

\appendix

\section{Effective wavelengths and reddening coefficients}
\label{sec:appendix}

To ease future fitting (and plotting), we list in \Tref{photometry}
the wavelengths at which one should evaluate the flux for spectra
which are close to Rayleigh-Jeans, as well as relative
reddening coefficients.  These were calculated as follows
\begin{eqnarray}
\langle\lambda\rangle&=&\left(
\frac{\int\lambda^{-4}\lambda R_\lambda\,{\rm d}\lambda}
     {\int\lambda R_\lambda\,{\rm d}\lambda}\right)^{-1/4},
\label{eq:l4eff}\\
\left\langle\frac{A_\lambda}{A_V}\right\rangle&=&
\frac{-2.5}{0.12}\log
\frac{\int 10^{-0.4\times0.12(A_\lambda/A_V)}
      \lambda^{-4}\lambda R_\lambda\,{\rm d}\lambda}
{\int \lambda^{-4}\lambda R_\lambda\,{\rm d}\lambda}.
\label{eq:aleff}
\end{eqnarray}
Note that the wavelength $\langle\lambda\rangle$ is not the effective
wavelength of the detected photons in a given filter, but is the
wavelength at which a $\lambda^{-4}$ spectrum has the same flux as a
$\lambda^0$ spectrum that produces the same number of counts in the
filter considered.

\bibliographystyle{apj}

\end{document}